\documentclass[preprint,showpacs,aps]{revtex4}
\usepackage{amssymb,amsmath,amsfonts,graphicx,latexsym,hyperref}

\begin{document}

\title{Coherent control of microwave pulse storage in superconducting circuits}

\author{Patrick M. Leung}
\author{Barry C. Sanders}
\affiliation{Institute for Quantum
Information Science, University of Calgary, Alberta T2N 1N4, Canada}

\begin{abstract}
Coherent pulse control for quantum memory is viable in the optical
domain but nascent in microwave quantum circuits. We show how to
realize coherent storage and on-demand pulse retrieval entirely
within a superconducting circuit by exploiting and extending
existing electromagnetically induced transparency technology in
superconducting quantum circuits. Our scheme employs a linear array
of superconducting artificial atoms coupled to a microwave
transmission line.
\end{abstract}

\pacs{42.50.Gy, 03.67.-a, 85.25.Hv}

\maketitle

Circuit quantum electrodynamics has successfully realized numerous
quantum optical phenomena in the microwave ($\mu$w)
domain~\cite{BFB+09} and is a promising architecture for quantum
information technology~\cite{ZB08}. Various superconducting
artificial atoms (SAAs) have been realized including Cooper pair
boxes~\cite{NPT}, transmons~\cite{KYG+99}, flux qubits~\cite{Chi03}
and fluxonium~\cite{Man09}. These SAAs can be coupled to a $\mu$w
transmission line that serves as a `quantum bus' for $\mu$w
photons~\cite{MCG+07}. For efficient large-scale quantum information
processing, synchronization of information flow between highly
separated SAAs requires the $\mu$w photons to be stored in quantum
memories~\cite{LST09} coherently and retrieved on-demand.

One strategy is to store photons in high quality superconducting
microwave resonators~\cite{MNB+12}, but they yield low
fidelities~\cite{MWY+11}. Another strategy is to couple the
superconducting circuit to external spin-$\frac{1}{2}$
systems~\cite{RDD+06} so that quantum information stored in the
$\mu$w field can be transferred for on-demand storage and
retrieval~\cite{Wu}. Exciting progress has been made in this
direction by coupling the superconducting circuit to nitrogen
vacancies in diamond~\cite{KOB+10} but the storage time is still
limited to $\sim30$~ns~\cite{KDD+12}, and hybridizing with another
quantum information medium adds extra complexity to the fabrication
process. Pulse storage may be achievable by controlling the
transition frequencies of SAAs~\cite{SPS+07} but this scheme lacks
coherent control for quantum memory. We propose to create a linear
array of SAAs coupled to the transmission line, with this array
serving as a pulse storage and on-demand retrieval medium. Our
scheme is similar to quantum memory approaches in atomic
systems~\cite{BZL03} and leverages off recent experimental
demonstrations of superconducting-circuit electromagnetically
induced transparency (EIT)~\cite{AAZ+10}. The point of our scheme is
to create $\mu$w quantum memory in a superconducting circuit without
the need to hybridize media, akin to the optical case with atomic
quantum memory~\cite{LST09}.

\begin{figure}
\centerline{\includegraphics[height=0.8cm]{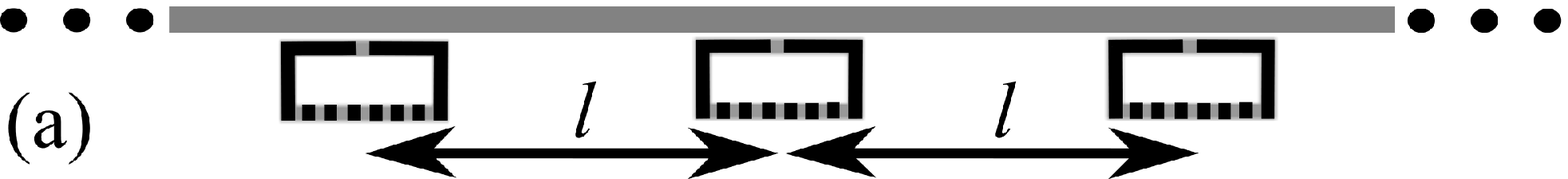}}
\vspace{0.3cm}
\includegraphics[height=2.5cm]{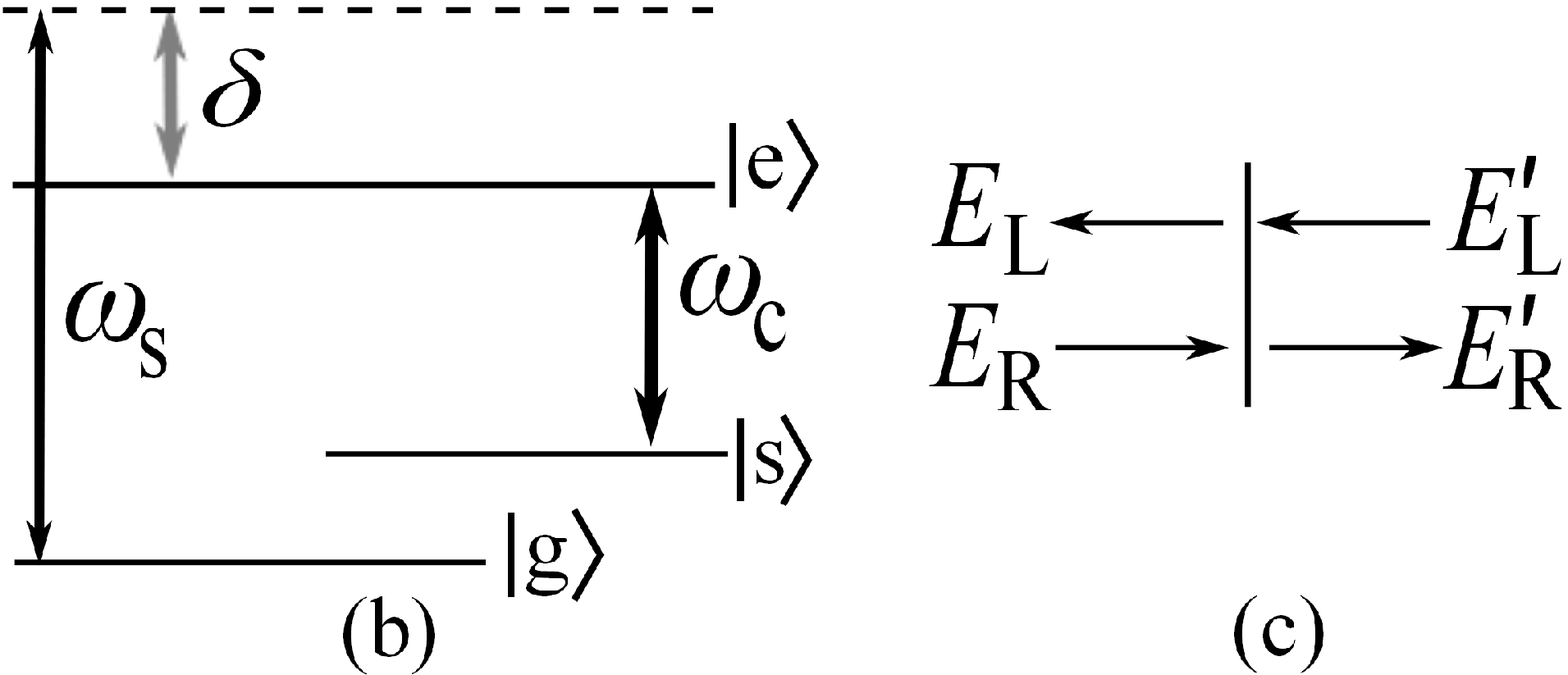}
\caption{
    (a)~Linear array of fluxonium (rectangles with dashed side)
    periodically spaced with separation~$l$ and
    coupled to (gray) transmission line.
    (b)~Three energy levels of the fluxonium with on-resonant control driving of the~$|\text{e}\rangle\leftrightarrow|\text{s}\rangle$
    transition with frequency~$\omega_\text{c}$ and off-resonant driving of the~$|\text{e}\rangle\leftrightarrow|\text{g}\rangle$ transition with
    frequency~$\omega_\textrm{s}$ and detuning~$\delta$.
    (c)~Incoming and outgoing fields (indicated by arrow direction) for one SAA labeled by field amplitudes $E_\text{R}$ and $E'_\text{L}$ for
    incoming fields and $E_\text{L}$ and $E'_\text{R}$ for outgoing, with~L
    and~R denoting `left' and `right', respectively.} \label{fig:array}
\end{figure}

Although various choices of SAAs are available, we explore using
fluxoniums~\cite{Man09} because they can serve as highly-coherent
three-level SAAs~\cite{MMK+12}. Three-level fluxoniums are
particularly amenable to $\mu$w-field control via Autler-Townes
splitting or EIT~\cite{ADS11}. As illustrated in
Fig.~\ref{fig:array}(a), in our model, $n$ SAAs are periodically
spaced with inter-SAA distance~$l$ optimized to yield maximum
storage efficiency. The total medium length is $d=(n-1)l$. Aperiodic
spacings affect the position of the scattering resonances in the EIT
spectra but do not alter the efficacy of our scheme. We shall see
that interference due to inter-SAA scattering modifies the
transmission spectrum in a geometry-dependent way, but these complex
spectral features are not an obstacle to realizing $\mu$w pulse
storage and retrieval.

The energy levels $|\text{g}\rangle$, $|\text{s} \rangle $ and
$|\text{e} \rangle $ and driving fields for each three-level SAA are
depicted in Fig.~\ref{fig:array}(b). Two driving fields are required
to realize EIT. The control-field frequency~$\omega_\text{c}$ is
resonant with $\omega_\text{es}$ (with $\omega_\text{xy}$ the
frequency for transition
$|\text{x}\rangle\leftrightarrow|\text{y}\rangle$ and
x,y~$\in\{\text{g},\text{s},\text{e}\}$) and a signal field of
frequency~$\omega_\textrm{s}$ detuned
by~$\delta=\omega_\textrm{s}-\omega_\text{eg}$ from
$\omega_\text{eg}$. When the control field is off, the SAA is opaque
to the signal field; turning on the control field makes the SAA
transparent. In addition, this three-level SAA can yield `EIT with
amplification'~\cite{JBBS10}.

Previous studies focusing on EIT in superconducting circuits
concentrate on the case of a single SAA and predict~\cite{MDO+04}
and demonstrate~\cite{AAZ+10} changes to the transmission and
reflection characteristics of the single SAA but do not consider
phenomena such as slow light and properties such as optical depth
that are crucial for building quantum memory. Here we consider a
linear array of $n$ SAAs to enable on-demand storage and retrieval
of $\mu$w pulses. Before the $\mu$w signal pulse arrives, a
large-amplitude long-duration $\mu$w control field coherently pumps
the medium from $|\text{s}\rangle$ to $|\text{e}\rangle$ via the
transmission line, and the field addresses the SAAs equally when
they are relaxed to $|\text{g}\rangle$. The pumping results in
coherence between the SAA energy levels, which suppresses the
absorption of the small-amplitude signal pulse as well as slowing
the signal propagation for pulse storage.

When the signal pulse has entered the SAA medium, the control field
is turned off to absorb the photons, which are converted to
dark-state polaritons for storage. The control field is subsequently
turned on to retrieve the signal pulse on-demand. This method of
$\mu$w pulse storage in a quantum circuit is inspired by the
technique for storing an optical pulse in an EIT-enabled
medium~\cite{FIM05}, which means the artificial atomic medium
requires large optical depth for high storage
efficiency~\cite{LST09}, and we show that our array of SAAs is
capable of delivering the requisite optical depth.

As the medium is one-dimensional (1D), the incoming signal fields
encountering a SAA are scattered into transmitted and reflected
components as depicted in Fig.~\ref{fig:array}(c). For a strong
non-depleting control field with constant Rabi frequency
modulus~$|\Omega|$ throughout the medium, the reflection coefficient
$r(\delta)$ for the signal field scattering from a SAA
is~\cite{AZA+10}
\begin{equation}
    r=\frac{\Gamma_{\text{eg}}}{2(\Gamma_{\text{e}}-i\delta)
        +\frac{|\Omega|^2}{2(\Gamma_\text{s}-i\delta)}},
    \Gamma_\text{e}=\frac{\Gamma_{\text{eg}}+\Gamma_{\text{es}}}{2},
    \Gamma_\text{s}=\frac{\Gamma_{\text{sg}}}{2},
\end{equation}
with~$\Gamma_\text{xy}$ the decay rate for the
$|\text{x}\rangle\leftrightarrow|\text{y}\rangle$ transition. By treating the SAAs as point
scatterers with linear susceptibilities, each SAA acts as a boundary
for the transmission and reflection components of the signal field.
Therefore, we apply the reflection-and-transmission transfer matrix
method~\cite{Yeh88} to calculate the transmission through a sequence
of SAAs. The transfer matrix method has been employed for studying analogs of EIT in
plasmonic systems but without considering pulse
storage~\cite{LLM12}. For a single SAA, the induced transformation
of the signal field is
\begin{equation}
    (1-r(\delta))\begin{pmatrix}E_\text{R}(\delta)\\E_\text{L}(\delta)\end{pmatrix}
    =\begin{pmatrix}1&-r(\delta)\\r(\delta)&1-2r(\delta)\end{pmatrix}
        \begin{pmatrix}E'_\text{R}(\delta)\\E'_\text{L}(\delta)\end{pmatrix}.
\label{eq:TransM}
\end{equation}
We include phase shift $\varphi=l\omega_\textrm{s}/c$, taking into
account the free propagation of the signal field between the SAAs
through the transmission line at speed $c$. The overall
transformation by the linear array, denoted as matrix $M$, is given
by
\begin{equation}
    \begin{pmatrix}E_{\text{1R}}(\delta)\\E_{\text{1L}}(\delta)\end{pmatrix}
        =M(\delta)\begin{pmatrix}E'_{n\text{R}}(\delta)\\E'_{n\text{L}}(\delta)\end{pmatrix}
\end{equation}
for the subscript~1 denoting the first SAA and~$n$ the last SAA of
the $n$-SAA array.

As $E_\text{in}:=E_{\text{1R}}$ is the only nonzero input field,
$E'_{n\text{L}}=0$, only the upper-left element of the response
matrix
\begin{align}
    M_{11}=&\frac{(A_{+}+B)(A_{-}+B)^n-(A_{+}-B)(A_{-}-B)^n}{2^{n+1}\text{e}^{i(n-1)\varphi}B(r-1)^n},
        \nonumber   \\
    A_{\pm}=&\pm\text{e}^{2i\varphi}(1-2r)-1,
        \nonumber   \\
    B=&\sqrt{(\text{e}^{2i\varphi}-1)(\text{e}^{2i\varphi}(1-2r)^2-1)}
    \label{eq:MAB}
\end{align}
is needed to obtain the output signal field
$E_\text{out}=E_\text{in}/M_{11}$, with transmission
$T=|1/M_{11}|^2$.

Let us now consider an example of how this system works by
judiciously choosing favorable yet realistic parameters. The
transition frequency values are $\omega_\text{eg}/2\pi=10.4$~GHz and
$\omega_\text{es}/2\pi=6.99$~GHz at flux bias $\Phi=0.32\Phi_0$
for~$\Phi_0\equiv h/2e$ the flux quantum; this flux bias yields long
coherence times~\cite{MMK+12}. The
$|\textrm{e}\rangle\leftrightarrow|\textrm{g}\rangle$ transition
wavelength is $\lambda=2\pi c/\omega_\text{eg}=11.6$~mm and
$c=1.2\times10^8$~m/s~\cite{WJP+11}. We treat decoherence as being
due fundamentally to spontaneous emission~\cite{AZA+10}. Using
Manucharyan et~al.'s~\cite{MMK+12} measured relaxation parameter
$\Gamma_\text{sg}\approx0.167$~MHz, we calculate the fluxonium
relaxation parameters to be $\Gamma_\text{eg}=173\Gamma_\text{sg}$
and $\Gamma_\text{es}=40\Gamma_\text{sg}$~\footnote{The relaxation
parameters are found using the transition matrix element relation
$\mu_\text{es}=0.23\mu_\text{eg}$ from Ref.~\cite{JBBS10}.}.

\begin{figure}
\includegraphics[height=7.0cm,angle=0]{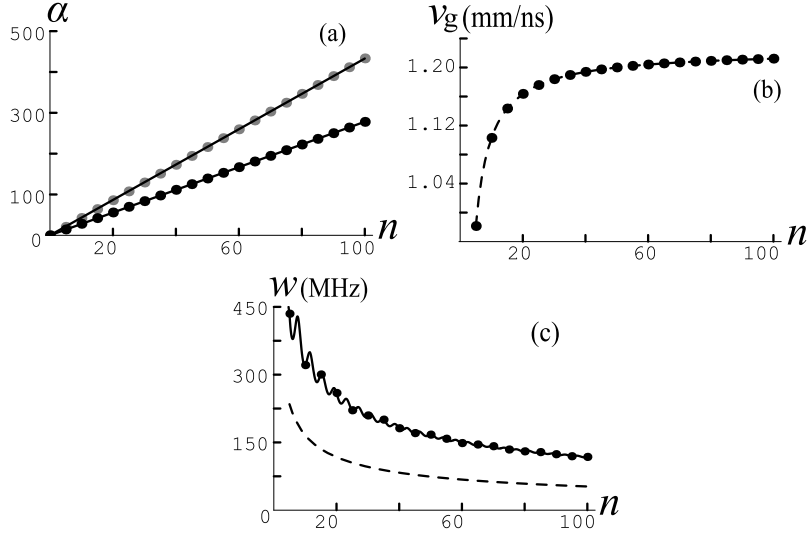}
\caption{
    Characteristics of a periodically spaced SAA array with~$n$ fluxoniums and inter-SAA separation~$l$.
    (a) Optical depth~$\alpha$ vs~$n$ for $l=\lambda/4=2.90$~mm (gray dots)
    fitted to a straight line $\alpha=4.33n$ and for $l=0.74$~mm (black dots) fitted to straight line $\alpha=2.78n$.
    For (b-c) the relevant parameters are $l=1.50$~mm and the control-field Rabi frequency is $\Omega=218$~MHz
    with numerical evaluation (dots) and the optical-EIT formulae (dashed).
    (b) Group velocity of a pulse through the array.
    (c) EIT window width~$w$ vs~$n$. Analytical evaluation (solid)
    of $w$ using Eq.~(\ref{eq:wnum}) shows oscillatory behavior.
    }
\label{fig:OpVgWTBR}
\end{figure}

The field transmission in a translucent medium is related to the
optical depth $\alpha$ by $T=\text{e}^{-\alpha}$. For equally spaced
SAAs, the length of the medium increases as $n$ increases and Beer's
Law holds if $\alpha\propto n$. For the choice of parameters and for
two choices of inter-SAA separation~$l$, at $\lambda/4$ and at one
other value, we can see in Fig.~\ref{fig:OpVgWTBR}(a) that the
optical depth at $\delta=0$ with control field off ($\Omega=0$) is a
linear function of~$n$ in accordance with Beer's Law; additional
modulation due to inter-SAA scattering interference is negligible.
For even a dozen fluxoniums the optical depth is high compared to
typical optical depths in optical-EIT of less than 25~\cite{PGN08}.
From equation~(\ref{eq:MAB}), we find that the medium absorbs the
signal field maximally when $l\to(2m+1)\lambda/4$, where $m$ is any
non-negative integer. An exception to Beer's Law occurs when $l\to
m\lambda/2$, in which
$\alpha\to2\textrm{ln}\left(\frac{1+(n-1)r(0)}{1-r(0)}\right)$.

\begin{figure}
\includegraphics[height=3.5cm]{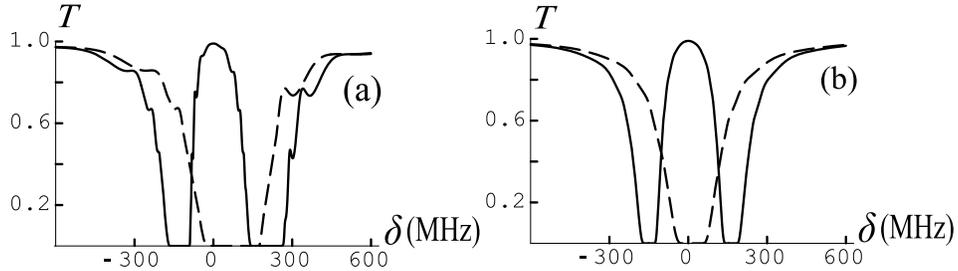}
\caption{
    Transmission through $n=100$ fluxoniums with inter-SAA separation (a) $l=0.3$~mm and (b) $l=1$~mm
    for $\Omega=0$ (dashed) and with $\Omega=309$~MHz (solid).
    }
\label{fig:BraggEIT}
\end{figure}

Inter-SAA scattering interference occurs during transmission of a
$\mu$w field through a periodic array of SAAs, which may have
complications that are absent in optical-EIT. We explore the effect
that inter-SAA scattering interference has on EIT within the 1D
array of fluxoniums as depicted in Figs.~\ref{fig:BraggEIT}(a)
and~\ref{fig:BraggEIT}(b).
Due to interference effects, in Fig.~\ref{fig:BraggEIT}(a) the
transmission profiles with and without the control field on are
markedly different from Fig.~\ref{fig:BraggEIT}(b).
Fig.~\ref{fig:BraggEIT}(b) is plotted using identical conditions as
for Fig.~\ref{fig:BraggEIT}(a) except with a different inter-SAA
spacing. In Fig.~\ref{fig:BraggEIT}(a) scattering creates asymmetry
and oscillations but the EIT features remain intact.
Fig.~\ref{fig:BraggEIT}(b) looks like textbook EIT, with $1-T$
replacing absorption, which is realized by choosing some appropriate
inter-SAA spacings.

In our scheme, the $\mu$w input signal field is a pulse, and its
speed is quantified by the group velocity. The transmitted signal
pulse acquires a phase factor $\text{e}^{ikd}$ from propagation
through the medium, and the dispersion relation and group velocity
are
\begin{equation}
\label{eq:vgnum}
    k(\omega_\textrm{s})=\frac{1}{d}\text{arg}\left(\frac{1}{M_{11}(\omega_\textrm{s})}\right), \,
    v_\text{g}=\left(\frac{\text{d}k}{\text{d}\omega_\textrm{s}}\right)^{-1}_{\delta=0},
\end{equation}
respectively. Based on
$T=T_{\delta=0}\text{exp}(-\delta^2/w^2)$~\cite{Lambropoulos} with
$T_{\delta=0}\approx1$, which holds for small $\delta$ (i.e.\ in the
vicinity of the EIT window), the EIT window width~$w$ is given by
\begin{equation}
\label{eq:wnum}
    w^2=-2\left(\left. \frac{\text{d}^2\text{ln}T}{\text{d}\delta^2}\right|_{\delta=0}\right)^{-1}.
\end{equation}
Using Eqs.~(\ref{eq:vgnum}) and~(\ref{eq:wnum}), we determine the
inter-SAA scattering-included group velocity and EIT window width.
Furthermore, we determine the group velocity and the width for the
case of negligible inter-SAA scattering as in optical EIT. When EIT
occurs, within the EIT window, the reflection of signal due to each
SAA is small. Therefore, for the case of negligible inter-SAA
scattering, the overall transmission coefficient is
$M_{11}^{-1}=(1-r)^n\text{e}^{i(n-1)\varphi}$ and the resultant
transmission is $T\approx\text{e}^{-2n\text{Re}(r)}$. Hence we have
\begin{equation}
\label{eq:wan}
    w\approx\frac{\left|\Omega\right|^2}{4\Gamma_\text{eg}\beta\sqrt{2n}}, \,
    \beta=|\Omega|^2\sqrt{\frac{(\Gamma_\text{e}+2\Gamma_\text{s})|\Omega|^2-4\Gamma_\text{s}^3}{2\Gamma_\text{eg}(4\Gamma_\text{e}\Gamma_\text{s}+|\Omega|^2)^3}}.
\end{equation}
For $\chi$ the linear susceptibility, the overall transmission
coefficient is $M_{11}^{-1}=\text{e}^{i\omega_\textrm{s}\chi
d/(2c)}\text{e}^{i(n-1)\varphi}$~\cite{Lambropoulos} so
\begin{equation}
\label{eq:vgan}
    v_\text{g}=\left(\frac{1}{c}+\frac{2n\Gamma_\text{eg}}{(n-1)l|\Omega|^2}\right)^{-1}.
\end{equation}
Equations~(\ref{eq:wan}) and~(\ref{eq:vgan}) are the 1D version of
the EIT window width and group velocity expressions from optical-EIT
theory~\cite{Lambropoulos}, and we refer to these two equations as
optical-EIT formulae. We evaluate the scattering-included
$v_\text{g}$ and $w$ and compare them against optical-EIT formulae
in Figs.~\ref{fig:OpVgWTBR}(b-c), respectively, to see how closely
the optical-EIT formulae apply in the case of a $\mu$w pulse
propagating through a fluxonium array. The agreement between the
inter-SAA scattering-free theory and scattering-included theory
shows that, with the chosen parameters, inter-SAA scattering
interference has negligible effect on the group velocity.
Furthermore, for the case of including scattering, $w$ is larger
with $w\propto\frac{1}{\sqrt{n}}$ plus an additional modulation due
to inter-SAA scattering, which appears as oscillations that dampen
exponentially as $n$ increases~\footnote{See supplementary material
for the EIT window width modulation.}.

To show that storage of a $\mu$w pulse in the fluxonium array is
indeed possible using the EIT phenomenon, we consider an input pulse
whose spectral profile is Gaussian with respect to detuning~$\delta$
with $\delta=0$ corresponding to input frequency~$\omega_\text{eg}$.
The spectral function for the dimensionless input field is thus a
Gaussian function $E_\text{in}(\delta)$ with standard deviation
$\sigma$ and normalized field strength according to
$\int|E_\text{in}(\delta)|^2\text{d}\delta=1$. In our numerical
evaluation, we determine the pulse storage efficiency
$\eta(n,\Omega,l,\sigma)$, which is defined as the percentage of the
pulse trapped inside the medium as soon as the control field is
turned off.

We now employ the following conditions to evaluate the efficiency
under reasonable experimental conditions. The medium must be
sufficiently transparent when the pulse enters, we set $\Omega$ such
that $T_{\varphi\to 0}=0.99$ at $\delta=0$. We require that $v_\text{g}\ll c$ when the
control field is on, which is guaranteed by choosing $l$ such that
$v_\text{g}\approx c/100$ from Eq.~(\ref{eq:vgan}). Not only does
this condition ensure that the pulse is slowed inside the medium but
also assures that the control field can be switched off to make the
medium opaque before the pulse escapes. The pulse must fit inside
the EIT spectral window so we choose $\sigma$ such that
$\int|E_\text{out}(\delta)|^2\text{d}\delta\approx 0.98$ as a
benchmark. If the pulse is temporally longer than the available time
window $d/v_\text{g}$, only the central part of the pulse within
this time window is stored and the pulse storage efficiency is
reduced.

\begin{figure}
\includegraphics[height=7.5cm]{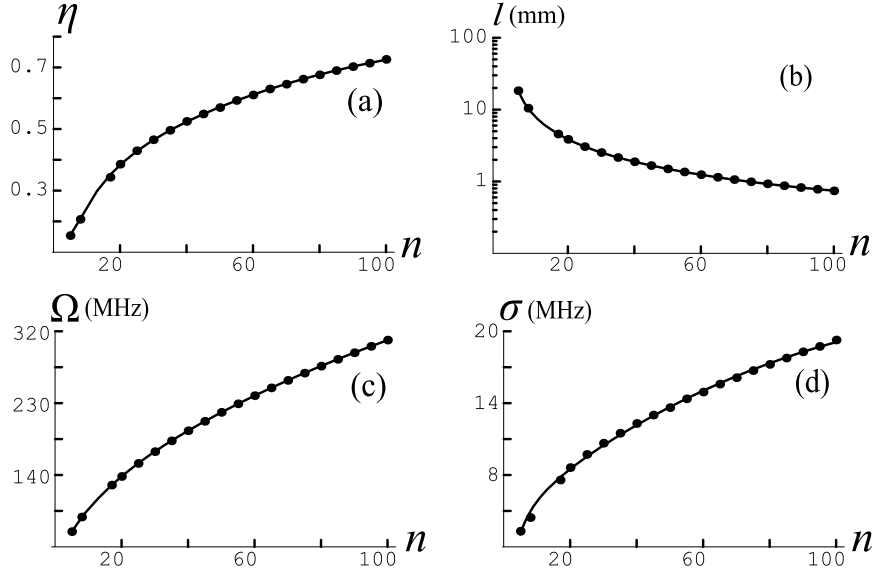}
\caption{
    (a) Optimal storage efficiency for $n$ fluxoniums. Solid line is
    an interpolated fit. (b) Corresponding fluxonium separation $l$. (c)
    Corresponding control-field Rabi frequency. (d) Corresponding
    standard deviation of the Gaussian spectral profile of the input pulse.
    }
\label{fig:EtaSepRabiSigma}
\end{figure}

Storage efficiency vs $n$ is shown in
Fig.~\ref{fig:EtaSepRabiSigma}(a). The corresponding values of SAA
separation $l$, control-field Rabi frequency $\Omega$ and the
standard deviation of the pulse $\sigma$ from the constraining
conditions are shown in Figs.~\ref{fig:EtaSepRabiSigma}(b-d)
respectively. Although some of the separation values in
Fig.~\ref{fig:EtaSepRabiSigma}(b) do lead to distortion in the EIT
spectra as in Fig.~\ref{fig:BraggEIT}(a), the effect on the
efficiency is negligible with the corresponding pulse widths. As the
number of SAAs increases, the pulse storage efficiency increases as
expected. For five fluxoniums, the efficiency is $\eta\approx0.15$,
and for $n=100$, the efficiency is $\eta\approx0.72$, which is
comparable to the (retrieval) efficiency observed in optical-EIT
pulse storage medium~\cite{PGN08}. Asymptotically, $\eta$ approaches
near unity as $n$ becomes large. For example we evaluate the
efficiency for $n=300$ and find that $\eta\approx0.91$. Therefore,
pulse storage is possible with a few fluxoniums and asymptotically
efficient with increasing number of fluxoniums. At $0.32\Phi_0$ flux
bias, the coherence time for fluxonium is about 1 $\mu$s~\cite{MMK+12}
so our proposed memory could store a pulse for approximately 1 $\mu$s with
negligible loss when EIT is off. Loss increases with the number of fluxoniums when EIT is on but strengthening $\Omega$ ameliorates this loss, see Fig. 4(c). The pulse is retrieved after storage by turning on the
control field and is optimally retrieved if
the retrieval pulse is the time-reversal of the input
pulse~\cite{PGN08}.

In conclusion, we have shown that coherent storage and on-demand
pulse retrieval are feasible using existing superconducting
artificial atom technology~\cite{Man09} provided that an array of
sufficiently large number of the artificial atoms is constructed as
suggested. Our scheme obviates the need to hybridize the technology,
e.g. by coupling to nitrogen-vacancy diamond defects~\cite{KOB+10}.\
Hybridizing is a promising approach but introduces unwanted
complexity and the storage time is limited by inhomogeneous
broadening. Our scheme is surprisingly efficient even for few
artificial atoms, for example delivering 15\% efficiency with just
five fluxoniums and with approximately 1 $\mu$s storage time. This
efficiency is dramatically improved with more fluxoniums and
ultimately reaches near unit efficiency asymptotically. Our
scattering-free 1D optical-EIT equations provide a good guideline
for how to construct quantum memory in superconducting circuits.
Numerical evaluations are then necessary to assure that the theory
gives quantitative agreement with experiments. Our results are
important for information synchronization and quantum memory in
superconducting circuits, which are some of the most promising
quantum computing technologies.

We appreciate financial support from AITF, NSERC and CIFAR, and
valuable discussions with A. Blais and J. Joo.

\section{Supplementary Material}
Inter-SAA scattering interference occurs in a one-dimensional array
of superconducting artificial atoms, which leads to oscillations in
the electromagnetically induced transparency window width $w$ as a
function of the number of fluxoniums $n$. Here we derive an
expression for $w(n)$ that shows the $w\propto\frac{1}{\sqrt{n}}$
behavior with an additional oscillatory modulation term. At the
vicinity of zero signal field detuning, i.e.\ $\delta\approx0$, the
transmission $T(\delta)=T(0)\textrm{exp}(-\frac{\delta^2}{w^2})$
with $T(0)\approx1$. Hence, the window width is
\begin{equation}
    w^2=\frac{2}{\left.-\frac{\text{d}^2\text{ln}T}{\text{d}\delta^2}\right|}_{\delta=0},
    \nonumber
\end{equation}
where the transmission $T=\left|\frac{1}{M_{11}}\right|^2$. For $r$
the reflection coefficient for a single fluxonium and $\phi$ the
phase shift due to free propagation between a pair of adjacent
fluxoniums,
\begin{align}
    M_{11}=&\sum_{j=1,2}G_jF_j^n,
        \nonumber   \\
    G_1=&\frac{A_++B}{2\textrm{e}^{-i\phi}B},
        \nonumber   \\
    G_2=&-\frac{A_+-B}{2\textrm{e}^{-i\phi}B},
        \nonumber   \\
    F_1=&\frac{A_-+B}{2\textrm{e}^{i\phi}(r-1)},
        \nonumber   \\
    F_2=&\frac{A_--B}{2\textrm{e}^{i\phi}(r-1)},
        \nonumber   \\
    A_{\pm}=&\pm\text{e}^{2i\varphi}(1-2r)-1,
        \nonumber   \\
    B=&\sqrt{(\text{e}^{2i\varphi}-1)(\text{e}^{2i\varphi}(1-2r)^2-1)}.
        \nonumber
\end{align}
Note that the square root in the expression for $B$ may take a plus
or minus sign. Changing the sign leads to a swap of expression for
$G_1$ and $G_2$ as well as for $F_1$ and $F_2$; thus, the expression
for $M_{11}$ is the same for either case. In the following, we
obtain an expression for
$-\frac{\text{d}^2\text{ln}T}{\text{d}\delta^2}$.
\begin{align}
-\textrm{ln}T=&\textrm{ln}\sum_{j=1,2}G_jF_j^n+\textrm{c.c.},
        \nonumber   \\
-\frac{\textrm{dln}T}{\textrm{d}\delta}=&\frac{1}{\sum_{j=1,2}G_jF_j^n}\sum_{j=1,2}\left(nG_j\frac{\textrm{dln}F_j}{\textrm{d}\delta}+\frac{\textrm{d}G_j}{\textrm{d}\delta}\right)F_j^n+\textrm{c.c.},
        \nonumber \\
-\frac{\textrm{d}^2\textrm{ln}T}{\textrm{d}\delta^2}=&\frac{1}{\sum_{j=1,2}G_jF_j^n}\sum_{j=1,2}\left(G_j\left(\frac{\textrm{dln}F_j}{\textrm{d}\delta}\right)^2n^2+\left(G_j\frac{\textrm{d}^2\textrm{ln}F_j}{\textrm{d}\delta^2}+2\frac{\textrm{d}G_j}{\textrm{d}\delta}\frac{\textrm{dln}F_j}{\textrm{d}\delta}\right)n+\frac{\textrm{d}^2G_j}{\textrm{d}\delta^2}\right)F_j^n
        \nonumber   \\
        &-\frac{1}{(\sum_{j=1,2}G_jF_j^n)^2}\left(\sum_{j=1,2}\left(nG_j\frac{\textrm{dln}F_j}{\textrm{d}\delta}+\frac{\textrm{d}G_j}{\textrm{d}\delta}\right)F_j^n\right)^2+\textrm{c.c.}
        \nonumber
\end{align}

We define
\begin{align}
a_j=&G_j\left(\frac{\textrm{dln}F_j}{\textrm{d}\delta}\right)^2,
    \nonumber   \\
b_j=&G_j\frac{\textrm{d}^2\textrm{ln}F_j}{\textrm{d}\delta^2}+2\frac{\textrm{d}G_j}{\textrm{d}\delta}\frac{\textrm{dln}F_j}{\textrm{d}\delta},
    \nonumber   \\
c_j=&\frac{\textrm{d}^2G_j}{\textrm{d}\delta^2},
    \nonumber   \\
b'_j=&G_j\frac{\textrm{dln}F_j}{\textrm{d}\delta},
    \nonumber   \\
c'_j=&\frac{\textrm{d}G_j}{\textrm{d}\delta}.
    \nonumber
\end{align}
such that
\begin{align}
-\frac{\textrm{d}^2\textrm{ln}T}{\textrm{d}\delta^2}=&\frac{(a_1n^2+b_1n+c_1)F_1^n+(a_2n^2+b_2n+c_2)F_2^n}{G_1F_1^n+G_2F_2^n}-\left(\frac{(b'_1n+c'_1)F_1^n+(b'_2n+c'_2)F_2^n}{G_1F_1^n+G_2F_2^n}\right)^2+\textrm{c.c.}
    \nonumber   \\
    =&\frac{1}{(G_1F_1^n+G_2F_2^n)^2}\Big{(}\left((G_1F_1^n+G_2F_2^n)(a_1F_1^n+a_2F_2^n)-(b'_1F_1^n+b'_2F_2^n)^2\right)n^2
    \nonumber   \\
    &+\left((G_1F_1^n+G_2F_2^n)(b_1F_1^n+b_2F_2^n)-2(b'_1F_1^n+b'_2F_2^n)(c'_1F_1^n+c'_2F_2^n)\right)n
    \nonumber   \\
    &+(G_1F_1^n+G_2F_2^n)(c_1F_1^n+c_2F_2^n)-(c'_1F_1^n+c'_2F_2^n)^2\Big{)}+\textrm{c.c.}
    \nonumber   \\
    =&2\textrm{Re}\Big{(}\frac{1}{(G_1F_1^n+G_2F_2^n)^2}\Big{(}\left(a_2G_1+a_1G_2-2b'_1b'_2\right)n^2
    \nonumber   \\
    &+\left((b_1G_1-2b'_1c'_1)F_1^{2n}+(b_2G_1+b_1G_2-2b'_1c'_2-2b'_2c'_1)+(b_2G_2-2b'_2c'_2)F_2^{2n}\right)n
    \nonumber   \\
    &+(c_1G_1-{c'_1}^2)F_1^{2n}+(c_2G_1+c_1G_2-2c'_1c'_2)+(c_2G_2-{c'_2}^2)F_2^{2n}\Big{)}\Big{)}
    \nonumber
\end{align}
as $F_1F_2=1$, $a_1G_1={b'_1}^2$ and $a_2G_2={b'_2}^2$.
For the chosen parameters (see paper), at $\delta=0$, we have
$F_1=0.689+i0.725$, $F_2=0.689-i0.725$,
$G_1=3.33\times10^{-9}+i3.50\times10^{-9}$, $G_2=0.689+i0.725$ and
$F_1-G_2=-6.94\times10^{-5}-i7.30\times10^{-5}$; thus,
$|G_1||F_1|^n\ll|G_2||F_2|^n$ for all $n$. Also,
$a_2G_1+a_1G_2-2b'_1b'_2=1.46\times10^{-27}-i2.88\times10^{-26}$,
$b_1G_1-2b'_1c'_1=3.32\times10^{-35}-i1.49\times10^{-35}$,
$b_2G_1+b_1G_2-2b'_1c'_2-2b'_2c'_1=-2.87\times10^{-23}+i5.66\times10^{-22}$,
$b_2G_2-2b'_2c'_2=-1.42\times10^{-18}+i6.38\times10^{-19}$,
$c_1G_1-{c'_1}^2=-3.40\times10^{-28}+i6.71\times10^{-27}$,
$c_2G_1+c_1G_2-2c'_1c'_2=7.01\times10^{-20}-i1.39\times10^{-18}$ and
$c_2G_2-{c'_2}^2=-7.01\times10^{-20}+i1.39\times10^{-18}$.
Therefore, the $n^2$ term and the $F_1^{2n}$ terms are
insignificant. The simplified expression for the window width in Hz
is
\begin{equation}
w\approx\frac{1}{\sqrt{7.10\times10^{-19}n+1.39\times10^{-18}(1-(0.9999)^{2n}\textrm{cos}1.62n)}}.
\end{equation}
\end{document}